# Spin-Transfer-Torque Induced Spatially Nonuniform Switching in Ferrimagnets

Xue Zhang[1,2,3], Zhengde Xu[1,2,3], Jie Ren[1], Yixiao Qiao[1], Weijia Fan[4†] and Zhifeng Zhu[1,5†]

[1]School of Information Science and Technology, ShanghaiTech University, Shanghai, China 201210

[2]Shanghai Institute of Microsystem and Information Technology, Chinese Academy of Sciences, Shanghai, China, 200050

[3]University of Chinese Academy of Sciences, Beijing, China, 100049

[4]Shanghai Key Laboratory of Special Artificial Microstructure Materials and Technology and School of Physics Science and Engineering, Tongji University, Shanghai, China 200092

[5]Shanghai Engineering Research Center of Energy Efficient and Custom AI IC, Shanghai, China 201210

**Abstract**

Ferrimagnet (FiM), $(FeCo)_{1-x}Gd_x$, attracts research attention due to its ultrafast magnetic dynamics and finite net magnetization. Incorporating FiM into the magnetic tunnel junction will be beneficial to further improve the writing speed of magnetic random access memory (MRAM). It is commonly assumed that the FeCo and Gd atoms are switched together due to the strong exchange coupling, which remains valid even if one performs the two-sublattice macrospin simulation. Interestingly, using the atomistic model developed by our group, it is clearly seen that different atoms are not switched together. In addition, our study reveals that the nature of switching is spatially nonuniform even in the small sample with the dimension of 20 nm×20 nm. Furthermore, the characteristics of nonuniformity are completely different for samples with different Gd composition ($x$). When $x$ is close to the magnetization compensation point, successful switching cannot be obtained, but is accompanied by the stable oscillation.

The atom type that dominates the oscillation is different from that predicted by the two-sublattice macrospin model. In addition, the size of singular region is a non-monotonic function of current density. All these results can only be understood by considering the spatial nonuniform magnetization dynamics.

As an important component of magnetic random access memory (MRAM)[1-8], spin-transfer-torque magnetic tunnel junction (STT-MTJ)[9-13] has been widely studied in the past decade. The vertically applied current switches the magnetic moments in the free layer, resulting in the change of the stored information. To seek a more effective device with a lower writing current, the perpendicular MTJ (pMTJ) is proposed[14-16]. For the conventional ferromagnet (FM) based pMTJ, the frequency of the magnetization dynamics is in the Gigahertz (GHz) range[17]. To speed up the switching, researchers turned to the materials which possess a strong exchange field. Recent studies about the antiferromagnet (AFM)[18-20] demonstrated that a stagger magnetic field can be used to excite magnetization switching[21]. However, the efficient manipulation and readout of magnetic states in AFM remain a challenge.

In contrast to the AFM, the ferrimagnet (FiM) [22-31] which consists of two different sublattices exhibits finite magnetization. This allows direct control using the current or field. In addition, easy detection using the tunnel magnetoresistance effect (TMR) can be achieved. More importantly, due to the large exchange field, FiM can be switched within picoseconds[32-35]. Because of these properties, the FiM based pMTJ becomes a promising alternative to the FM based pMTJ.

Some works on the atomistic study of FiM switching have been carried out in the past few years[36-42]. This method has been proved useful in revealing finer details compared to the macrospin model and the micromagnetic model[40]. However, the STT induced switching in FiM is not well understood. In this paper, we have developed an atomistic model with the STT effect to study the magnetization dynamics in the FiM, $(FeCo)_{1-x}Gd_x$, where $x$ denotes the Gd composition. The switching is achieved in picosecond time scale, which is much faster than the FM. More importantly, we find that the switching of the sample is nonuniform[43], and the switched region expands via domain wall (DW) motion. When $x < x_{MC}$ (magnetization compensation), the reversal tends to start first in the FeCo-rich[44-46] region at the corner. When $x$ is increased, i.e., $x$ approaches $x_{MC}$, a stable oscillation occurs with the appearance of singular regions. For an even larger $x$, i.e., $x > x_{MC}$, the switching first occurs in the Gd-rich region. In addition, the size of singular region also varies non-monotonically as a function of current density. We find that these switching and oscillation results can be explained only when we consider the atom distribution and atomic effective field. Since $x$ is frequently used as an important degree of freedom in tuning the FiM properties, our study contributes to the correct understanding of spin torque induced magnetization dynamics.

Figure 1(a) shows the device structure of FM/MgO/FiM MTJ with perpendicular magnetic moments. The vertically applied charge current density ($\mathbf{J}_c = \pm J_c \hat{z}$) produces STT[47]. The spin moment of the pinned layer ($\mathbf{S}_P$) is always along the +z direction. The magnetization dynamics of the free layer is studied by the atomistic model developed by ourselves[40, 48]. The Landau-Lifshitz-Gilbert-Slonczewski (LLGS) equations are numerically integrated through the fourth-order Runge-Kutta method. The sample is studied with open boundary conditions[49]. The schematic diagram of the model is shown in Fig. 1(b), where Gd and FeCo atoms are randomly distributed. The insert figure also shows that atoms are subjected to additional exchange interactions from the upper and lower layers.

The magnetization dynamics of every atom is described by the LLGS equation:

$$\frac{\partial \mathbf{S}_i}{\partial t} = -\gamma_i \mathbf{S}_i \times \mathbf{H}_{\text{eff},i} + \alpha \mathbf{S}_i \times \frac{\partial \mathbf{S}_i}{\partial t} - \gamma_i B_{D,i} \mathbf{S}_i \times (\mathbf{S}_i \times \mathbf{S}_p). \quad (1)$$

converting the LLGS equation into the Landau-Lifshitz (LL) form, we have:

$$\frac{\partial \mathbf{S}_i}{\partial t}(1+\alpha^2) = -\gamma_i \mathbf{S}_i \times \mathbf{H}_{\text{eff},i} - \gamma_i \alpha \mathbf{S}_i \times (\mathbf{S}_i \times \mathbf{H}_{\text{eff},i}) + \alpha\gamma_i B_{D,i} \mathbf{S}_i \times \mathbf{S}_p - \gamma_i B_{D,i} \mathbf{S}_i \times (\mathbf{S}_i \times \mathbf{S}_p) \quad (2)$$

where $i = 1,2$ denotes two different sublattices. $\gamma$ is the gyromagnetic ratio. $\mathbf{H}_{\text{eff}}$ is the effective field consisting of the exchange and anisotropy field. The specific expression of $\mathbf{H}_{\text{eff}}$ is obtained from the Hamiltonian $\mathcal{H} = A\sum_{i\neq j} \mathbf{S}_i \cdot \mathbf{S}_j - K\sum_i (\mathbf{S}_i \cdot \hat{\mathbf{z}})^2$, where $A$ is the exchange constant[35] with three different values, i.e., $A_{\text{Gd-Gd}} = -1.26 \times 10^{-21}$ J, $A_{\text{FeCo-FeCo}} = -2.83 \times 10^{-21}$ J, $A_{\text{FeCo-Gd}} = 1.09 \times 10^{-21}$ J. Here only the nearest neighbor interaction is considered. $K = 6.4080 \times 10^{-24}$ J is the uniaxial anisotropy constant[51]. The strength of STT is described as $B_{D,i} = \frac{\hbar}{2} \frac{J_c \eta d^3}{\mu_i e t_{FiM}}$, where $d = 0.4$ nm is the lattice constant, $\eta$ is the spin transfer efficiency which is equal to $0.8$[48], the saturation magnetization[42], $M_{s,\text{FeCo}} = 3.212 \times 10^5$ A/m, $M_{s,\text{Gd}} = 1.1056 \times 10^6$ A/m are related to the magnetic moment as $\mu_i/d^3$. In addition, the Hamiltonian of dipolar interaction is expressed as: $\mathcal{H}_{dipole} = -\frac{\mu_0}{4}\sum_{j\neq i} \frac{3(\mathbf{R}_{ij} \cdot \boldsymbol{\mu}_i)(\mathbf{R}_{ij} \cdot \boldsymbol{\mu}_j)}{R_{ij}^5} - \frac{\boldsymbol{\mu}_i \cdot \boldsymbol{\mu}_j}{R_{ij}^3}$. $\mathbf{R}_{ij}$ is the vector connecting $\boldsymbol{\mu}_i$ and $\boldsymbol{\mu}_j$. We have calculated the dipole field using the macro-cell method, and we find the dipole field has a negligible effect on the magnetization dynamics (Supplementary Note 1). We also ignore the stray field from the pinned layer since it is a constant field that is independent of $x$. All other parameters are the same as those in Ref. 51. Based on these parameters, the magnetization compensation $x_{MC} = 0.23$ and angular momentum compensation $x_{AMC} = 0.21$ are calculated by solving the equation $M_{s,FeCo} \times (1-x) = M_{s,Gd} \times x$ and $\frac{M_{s,FeCo}}{\gamma_{FeCo}} \times (1-x) = \frac{M_{s,Gd}}{\gamma_{Gd}} \times x$, respectively. We have also verified that the field-like torque (FLT) has negligible effect (Supplementary Note 2). In addition, from the LL equation, we find that the switching occurs only when the damping-like STT (the last term) overcomes the Gilbert damping (the second term on the right-hand side).

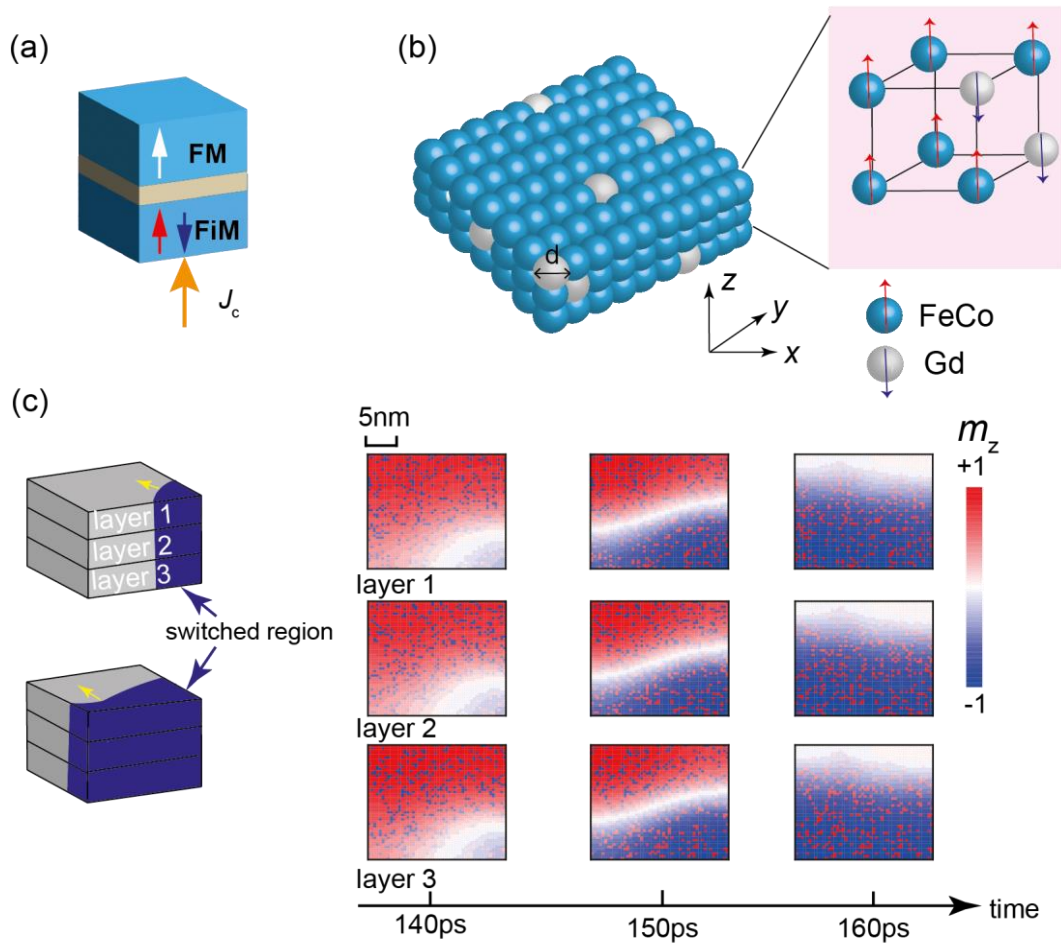

**Fig. 1.** Illustration of (a) the device structure, (b) the atomistic model, and (c) the layer-resolved nonuniform magnetization switching.

We first study the magnetization switching dynamics of a three-layer system (20 nm×20 nm×1.2 nm) with $\mathbf{J}_c = -1.5\times10^{11}$ A/m$^2$. The snapshot images of magnetization evolution in different layers are shown in Fig. 1(c). It clearly shows that the switching is spatial nonuniform in all layers. It is worth noting that although the atom distribution of each layer is different, their switching process is identical. In addition, we find the switching pattern of the three-layer system is similar to layer 1 (see Supplementary Note 3 for details). Therefore, we focus on the magnetization dynamics of layer 1.

Figure 2(a) shows the current driven averaged magnetization switching of FeCo atoms in samples (40nm×40nm×0.4nm) with different $x$. When $x = 0.15$, $m_{\text{FeCo},z}$ is switched from +1 to −1 under a negative $\mathbf{J}_c$ which points to the −z direction. The inset arrows in the red box depict this deterministic switching of the magnetic moments. However, when $x$ is further increased, i.e., $x = 0.2$, $m_{\text{FeCo},z}$ does not switch from +1 to −1. Therefore, we check the averaged magnetization dynamics of FeCo and Gd atoms which is shown in Fig. 2(b). We find that both atoms are oscillating stably in opposite directions which are drawn in the inset figure of Fig. 2(a) with a blue box. For an even larger $x$, i.e., $x = 0.5$, $m_{\text{FeCo},z}$ switches from +1 to −1 under a positive $\mathbf{J}_c$. This deterministic switching is illustrated in the inset figure with yellow box. Based on these results, we divide the above three cases into two groups, i.e., switching and oscillation. The switching group includes two samples of $x = 0.15$ and $x = 0.5$ while the oscillation group consists of the sample with $x = 0.2$.

To understand these magnetization reversal processes more in detail, we present snapshot images of the z-component of the magnetization of each atom in the sample at different times in Fig. 2(c). It can be seen that the switching process of different $x$ is significantly different. Furthermore, all samples exhibit nonuniform switching characteristics. For the switching group, when $x = 0.15$, the reversal starts with the formation of a switched region at the corner of the free layer at $t = 30$ ps. This switched region propagates via a DW, resulting in completed switching at $t = 70$ ps. However, when $x = 0.5$, the initial switching dynamics starts at the center of the free layer under a positive $J_c$. This reversed bubble-like region also expands through the DW to the entire sample, resulting in complete switching. For the oscillation group, the $m_z$ of the atoms at the corner changes first.

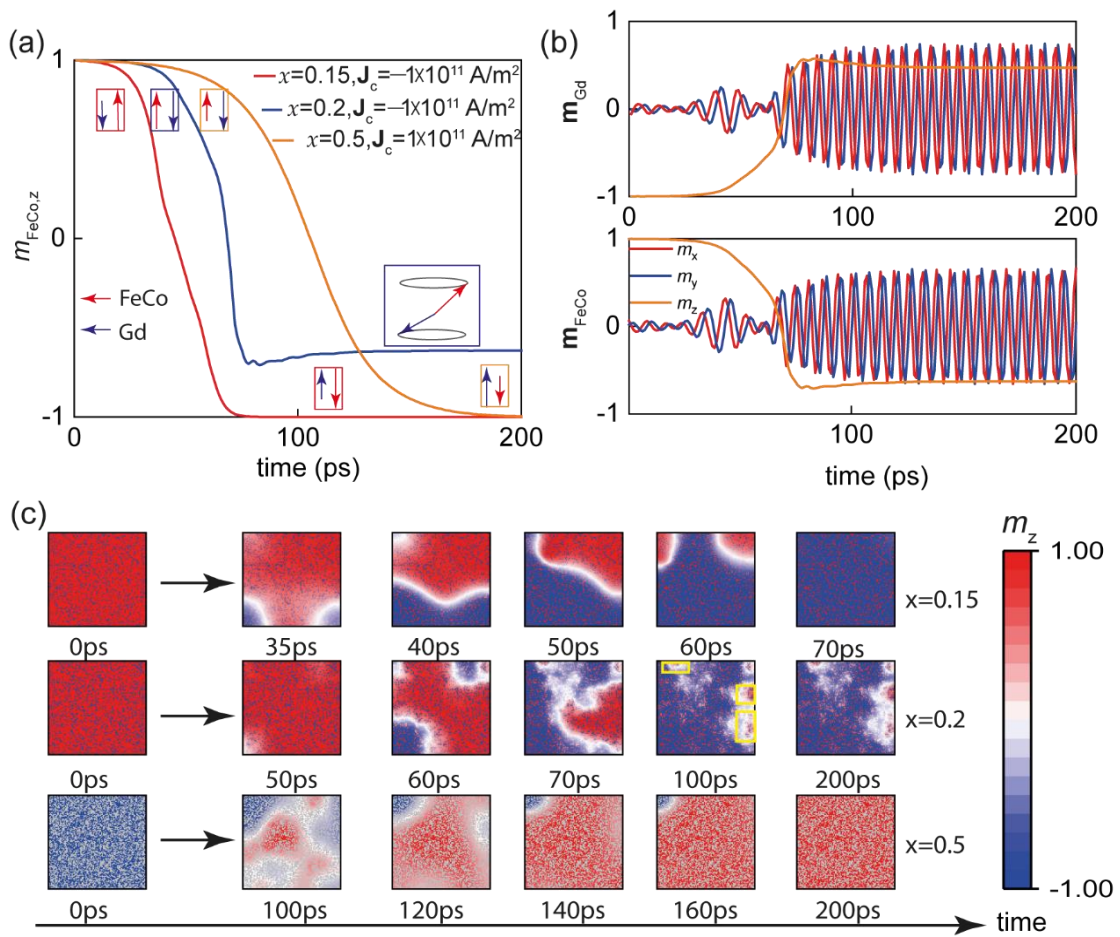

**Fig. 2.** (a) The averaged magnetization dynamics of samples with different $x$. (b) Magnetization dynamics of Gd and FeCo atoms when $x = 0.2$, $J_c = -1 \times 10^{11}$ A/m$^2$. (c) Spatial resolved

switching processes in different samples. In the sample with $x = 0.5$, the states of FeCo atoms are expressed in gray color for better visualization.

We noticed that the spatially nonuniform atom distribution, which is resulted from the random distribution of Gd atoms, coincides with the nonuniform switching process. Therefore, we check the atom distribution of the first switched region. For the switching group, the atom distribution of these two cases is shown in Fig. 3. The Gd-rich (FeCo-rich) region is defined as the region where the number of Gd atoms is larger (smaller) than 23 (i.e., $x_{MC} \times 100$), and they are represented in blue and red, respectively [see Fig. 3 and 4(a)]. The white area represents the composition region, in which the number of Gd atoms equals the composition of the whole sample, i.e., $x \times 100$. We define $\mathbf{m}_{net,Gd\text{-rich}}$, $\mathbf{m}_{net,FeCo\text{-rich}}$, $\mathbf{m}_{net,com}$ as the local net magnetization of the Gd-rich, FeCo-rich, and composition region, respectively. Differently, $\mathbf{m}_{net}$ represents the net magnetization of the entire sample. We find that when $x = 0.15$, the switching tends to start from the FeCo-rich region at the corner of the sample. However, when $x = 0.5$, the switching dynamics occurs first in the Gd-rich region. This corresponds to the results observed in Fig. 2(c).

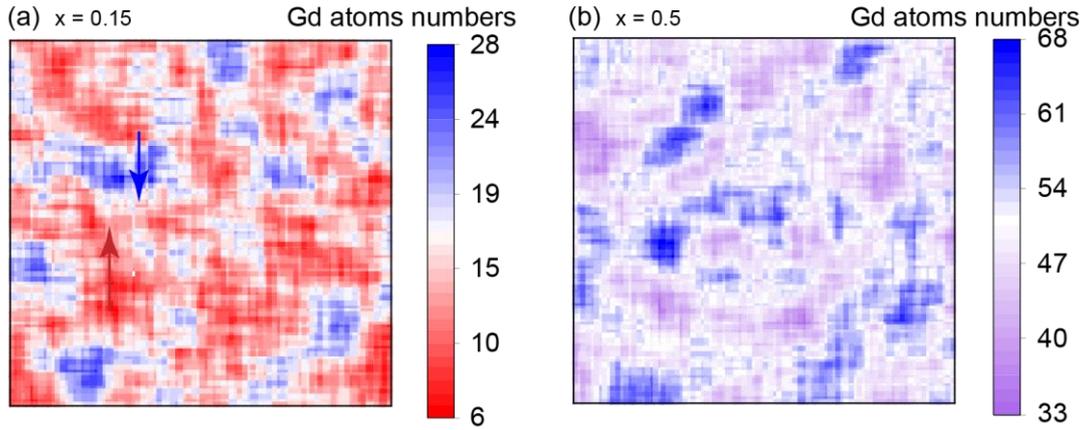

**Fig. 3.** Atom distribution in the sample with (a) $x = 0.15$ and (b) $x = 0.5$. The data in this figure represents the number of neighbor Gd atoms around a center atom. The red and blue areas represent the FeCo-rich and Gd-rich regions, respectively. See Supplementary Note 4 for the calculation details.

We attribute the observed nonuniform switching behavior to the different orientation of the local net magnetization. When $x < x_{MC}$, $\mathbf{m}_{net}$, $\mathbf{m}_{net,FeCo\text{-}rich}$, and $\mathbf{m}_{net,com}$ are in the same direction, i.e., +z direction. In contrast, $\mathbf{m}_{net,Gd\text{-}rich}$ points to –z direction due to the opposite alignment of Gd and FeCo atoms. Therefore, when $\mathbf{J}_c < 0$, the FeCo-rich region is first switched. Additionally, two factors contribute to the result that the reversal tends to occur at the corner of the sample. Firstly, the corner is a FeCo-rich region, which is first switched following the above discussions. Secondly, since the corner atoms have less neighbor atoms compared to the center atoms, the exchange field acting on them is smaller. Therefore, the damping-like STT [$-\mathbf{m}\times(\mathbf{m}\times\mathbf{S}_p)$] is easier to overcome the Gilbert damping torque [$-\mathbf{m}\times(\mathbf{m}\times\mathbf{H}_{eff})$], making them easier to reverse. We have also numerically verified that $\mathbf{H}_{ex}$ experienced by the atoms in the first switched region are smaller than those at the center.

Fig. 3(b) shows the atom distribution of the sample with $x > x_{MC}$ (i.e., $x = 0.5$), we find that the whole sample becomes Gd-rich region, i.e., there is no FeCo-rich region. In addition, **m**$_{net}$, **m**$_{net,Gd-rich}$, and **m**$_{net,com}$ are all pointing in −z direction. To switch this sample, one needs $\mathbf{J}_c > 0$. Since the switching occurs when the damping-like STT overcomes the Gilbert damping, the initial switched region can be determined by find out the local region with smallest **H**$_{eff}$. Note that the bule regions have more Gd-related exchange interactions which are smaller than FeCo-FeCo interaction, **H**$_{eff}$ in these regions will be smaller than the other regions. This explains the switching pattern shown in Fig. 2(c).

We have also studied the DW induced switching process. We find that during the DW propagation, the connected Gd (FeCo) atoms affect the DW shape when $x < x_{MC}$ ($x > x_{MC}$). This can be explained by noting the difference in the exchange constant between two sublattices. For example, when $x < x_{MC}$, the FeCo-rich region is switched by STT. However, the Gd-rich and averaged region will not be switched. Therefore, in these regions, the nature of the switching is the DW induced flip through the exchange interaction between sublattices. Notably, connected Gd clusters provide an easy path for the switching since the exchange interaction between Gd-Gd is larger than FeCo-Gd (Supplementary Note 5)[52].

In addition, the switched region expands via domain wall (DW) motion which is similar to the spin-orbit torque (SOT) driven magnetization behavior in Ref. 43 and 50. Previous studies also have shown that the different SOT strength on Gd and FeCo atoms, and the variation of exchange coupling between sublattices can also lead to the non-synchronized switching.

However, note that in the SOT switching, the easy axis of magnetization and the spin polarization are orthogonal, resulting in the requirement of an in-plane external field during the switching. This contrasts with the STT switching studied in this work, which can be directly analyzed by the torque balance condition.

For the oscillation sample, i.e., $x = 0.2$ which is close to $x_{MC}$, the snapshot images during the switching process are shown in the second row of Fig. 2(c). The color at the corner of the sample first changes from red to blue at 60ps when $\mathbf{J}_c < 0$. This blue region is expanded through the DW motion. However, the movement of DW does not turn the entire sample into blue. The appearance of the red and white region in the snapshot images after 100ps indicates that $m_z$ of atoms in different regions are different. Inspecting the atom resolved magnetization dynamics, we find that all atoms in the sample are oscillating stably. In addition, Gd atoms oscillate earlier than FeCo atoms, which is consistent with the conclusion in our previous work that Gd dominates the oscillation[53]. This is explained by analyzing the torque experienced by Gd and FeCo atoms using the two sublattices model, where the dominant atom is the one on which the damping-like STT and the Gilbert damping torque point in opposite directions and thus the oscillation is possible. We can then conclude that in the sample below $x_{MC}$, the oscillation is initiated by the Gd atom, the FeCo atom is then driven into oscillation due to the exchange interaction. However, in the sample with $x = 0.2$ shown in Fig. 2(c), the net magnetization of the Gd-rich region is in the same direction as $\mathbf{p}$, the above discussed Gd-induced oscillation condition cannot be satisfied following the two-sublattice model. This is in stark contrast to our results shown in Fig. 2(c) that oscillation is stable in the entire sample. Therefore, we define the abnormal oscillation regions as the singular regions. These contradictory results can only be explained by considering the nonuniform atom distribution in the sample, i.e., the oscillation first occurs in Gd atoms in the FeCo-rich region, which satisfy the oscillation condition. Afterwards, the atoms in Gd-rich region are dragged into oscillation due to the exchange interaction. The white regions appear when the Gd atoms precess to the x-y plane. In addition, the FeCo atoms in the Gd-rich region also satisfy the oscillation condition. They oscillate in the +z direction, resulting in the appearance of the red region. This explains the appearance of the singular regions.

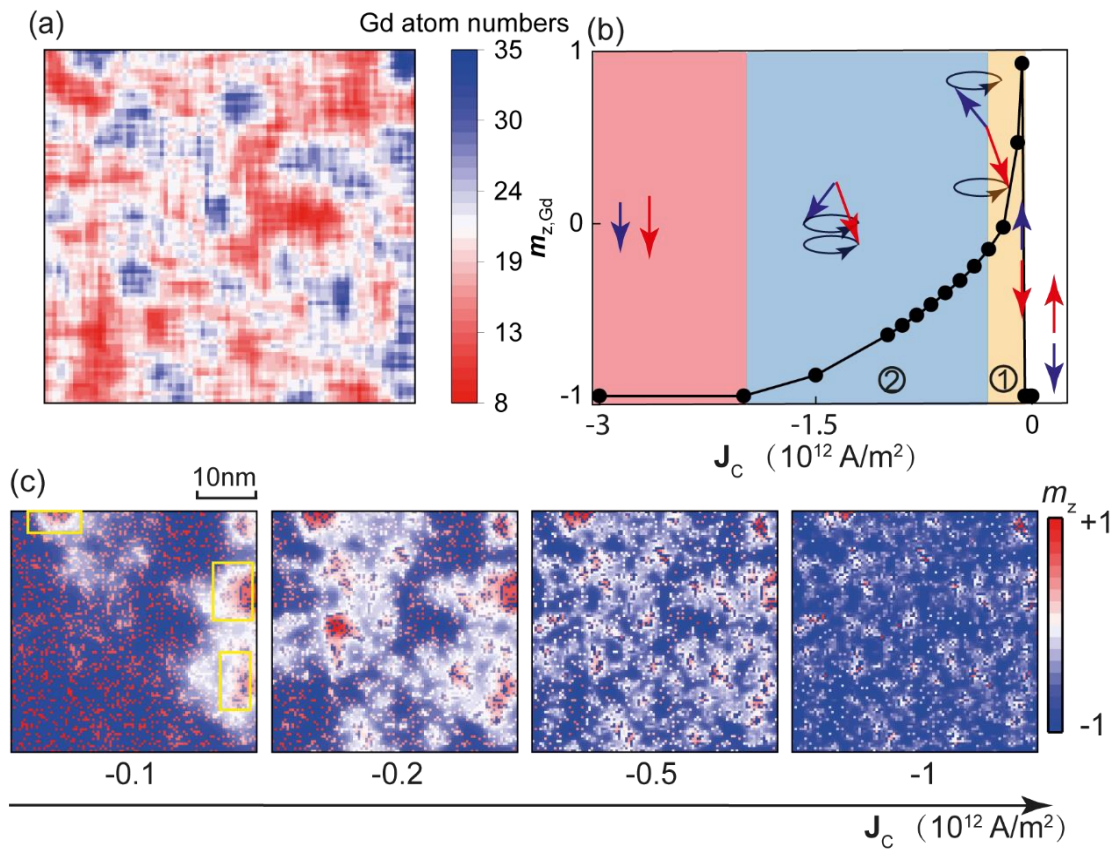

**Fig. 4.** (a) The atom distribution of the sample with $x = 0.2$. (b) Phase diagram of magnetization dynamics at $x = 0.2$. $m$ is calculated by averaging the atoms of the same type. (c) The snapshot images when the system is oscillating stably under different $J_c$.

Another result that cannot be explained in the two sublattice picture is the current induced oscillation profile. In Fig. 4(b), we illustrate the averaged magnetization phase diagram under different $\mathbf{J}_c$ for the sample with $x = 0.2$. As the net magnetization is sufficiently small and even zero at $x = x_{MC}$, it can be expected that the switching will not happen, and only the oscillation takes place. When $\mathbf{J}_c$ is increased, the oscillation changes from the exchange mode (region 1) to the flipped exchange mode (region 2)[53]. For an even large $\mathbf{J}_c$, all the magnetic moments are aligned to the –z direction. Therefore, in the oscillation region, when $\mathbf{J}_c$ is increased, the color of the entire sample should change to blue, i.e., enter the flip-exchange mode. However, as shown in Fig. 4(c), a completely blue snapshot does not occur even at very large $\mathbf{J}_c$, i.e., the singular regions are always present. In addition, when $\mathbf{J}_c$ is increased from $-0.1 \times 10^{12}$ A/m² to $-0.2 \times 10^{12}$ A/m², the number of singular regions increases. Interestingly, when $\mathbf{J}_c$ is further increased, the singular region shrinks to a very small area. For the contraction of the singular region, we consider two factors, i.e., the spin current and oscillation. Firstly, when $\mathbf{J}_c = -0.1 \times 10^{12}$ A/m², due to the large spin current, most Gd atoms in the white singular region are pulled to the –z direction. The decrease in the number of white regions leads to the contraction of the singular regions. In addition, since not all the Gd atoms are pointing to +z-axis even under a large $\mathbf{J}_c$, the Gd induced oscillation does not occur (Supplementary Note 6), and the precession of FeCo is responsible for the other oscillation regions. This requires that FeCo atoms are surrounded by more Gd atoms so that the FeCo atoms are always kept in the +z direction and received the correct torque. This picture is verified by the atom distribution in the singular region after shrinking, where only a few FeCo atoms in the Gd-rich region oscillate stably in the +z direction. We have analyzed the STT-induced oscillation using the two sublattice macrospin model in Ref. 53. Based on this model, it is expected that the oscillation in the sample with $x = 0.2$ will also be determined by the Gd atom under the two-sublattice model. However, in the atom-resolved magnetization state shown in Fig .4(c), we find that under a large $\mathbf{J}_c$, all the Gd atoms have been changed to the hemisphere with $m_{z,Gd} < 0$, whereas there are some FeCo atoms flipped to the +z direction. In this case, the oscillation condition, i.e., the balance between damping-like STT and Gilbert damping, can only be satisfied on these FeCo atoms with $m_{z,FeCo} > 0$. The appearance of these FeCo atoms with $m_{z,FeCo} > 0$ is induced by the formation of clustered atom regions, and we emphasize that these results cannot be explained using the macrospin model. Similar switching and oscillation results have been obtained in samples with different shapes (Supplementary Note 7). We want to emphasize that these results can only be explained by considering the spatially nonuniform magnetization dynamics.

In conclusion, we have developed an atomistic model to study the STT-induced magnetization switching in FiM. We find the switching is spatially nonuniform. The first switched region is determined by the dominant atom as well as the atomic effective field. In addition, a stable oscillation appears when $x$ approaches $x_{MC}$ with the appearance of singular regions. We find that the variation of the singular region can only be explained by considering the nonuniform atom distribution. It is thus concluded that compared to the macrospin model, the atomistic model used here is indispensable to correctly understand the magnetization dynamics of FiM.

## Supplementary Material

See the supplementary material for the dipole field, effect of FLT on the switching process, switching in the three-layer system, calculation details of Fig. 3, effect of the connected atoms on the switching dynamics, the Gd dominated oscillation, switching processes in a circular sample, and P to AP and AP to P switching.

We acknowledge the support from the National Key R&D Program of China (Grant No. 2022YFB4401700), National Natural Science Foundation of China (Grants Nos. 12104301 and 62074099) and Shanghai Sailing Program (Grant No. 20YF1430400).

## Data Availability

The data that support the findings of this study are available from the corresponding author upon reasonable request.

Corresponding Authors: †zhuzhf@shanghaitech.edu.cn; fanwj@tongji.edu.cn

# Spin-Transfer-Torque Induced Spatially Nonuniform Switching in Ferrimagnets

Xue Zhang[1,2,3], Zhengde Xu[1,2,3], Jie Ren[1], Yixiao Qiao[1], Weijia Fan[4†] and Zhifeng Zhu[1,5†]

[1]School of Information Science and Technology, ShanghaiTech University, Shanghai, China 201210

[2]Shanghai Institute of Microsystem and Information Technology, Chinese Academy of Sciences, Shanghai, China, 200050

[3]University of Chinese Academy of Sciences, Beijing, China, 100049

[4]Shanghai Key Laboratory of Special Artificial Microstructure Materials and Technology and School of Physics Science and Engineering, Tongji University, Shanghai, China 200092

[5]Shanghai Engineering Research Center of Energy Efficient and Custom AI IC, Shanghai, China 201210

## Supplemental Materials

**Note 1: Effect of the dipole field**

The Hamiltonian of the dipole is expressed as $\mathcal{H}_{dipole} = -\frac{\mu_0}{4\pi}\sum_{j\neq i}\frac{3(\boldsymbol{R}_{ij}\cdot\boldsymbol{\mu}_i)(\boldsymbol{R}_{ij}\cdot\boldsymbol{\mu}_j)}{R_{ij}^5} - \frac{\boldsymbol{\mu}_i\cdot\boldsymbol{\mu}_j}{R_{ij}^3}$, where $\boldsymbol{R}_{ij}$ is the vector connecting magnetic moments $\boldsymbol{\mu}_i$ and $\boldsymbol{\mu}_j$, the resulting dipole field is $\boldsymbol{H}_{dipole} = -\frac{\mu_0}{4\pi}\sum_{j\neq i}[\frac{3(\boldsymbol{R}_{ij})(\boldsymbol{R}_{ij}\cdot\boldsymbol{\mu}_j)}{R_{ij}^5} - \frac{\boldsymbol{\mu}_j}{R_{ij}^3}]$.

We then use this equation to calculate the dipole field in our sample with the size of 40 nm × 40 nm. Since the lattice constant is 0.4 nm, our sample consists of 10000 atoms, and one needs to compute the dipole field $10^8$ times per time step, which takes too long to get the complete magnetization dynamics. Instead, the macro-cell method is widely used to speed up the calculation, in which a group of atoms are treated as one macro-cell and then the above equation is used to calculate the dipole field between these macro-cells. The atoms in the same macro-cell are assumed to receive the same dipole field. This method is illustrated in Fig. S1, where we have 25 macro-cells for our sample. Note that the nonuniform switching happens in the whole film, the mesh grid of the macro-cell should be sufficient for the dipole field to manifest itself in the nonuniform switching.

Fig. S2 shows the comparison between the sample with and without the dipole field. It is clearly seen that both the switching and oscillation processes are not affected by the dipole field.

| $S_{av\,1,1}$ | $S_{av\,1,2}$ | $S_{av\,1,3}$ | $S_{av\,1,4}$ | $S_{av\,1,5}$ |
|---|---|---|---|---|
| $S_{av\,2,1}$ | $S_{av\,2,2}$ | $S_{av\,2,3}$ | $S_{av\,2,4}$ | $S_{av\,2,5}$ |
| $S_{av\,3,1}$ | $S_{av\,3,2}$ | $S_{av\,3,3}$ | $S_{av\,3,4}$ | $S_{av\,3,5}$ |
| $S_{av\,4,1}$ | $S_{av\,4,2}$ | $S_{av\,4,3}$ | $S_{av\,4,4}$ | $S_{av\,4,5}$ |
| $S_{av\,5,1}$ | $S_{av\,5,2}$ | $S_{av\,5,3}$ | $S_{av\,5,4}$ | $S_{av\,5,5}$ |

Fig. S1 Illustration of the macro-cell method in the calculation of dipole field.

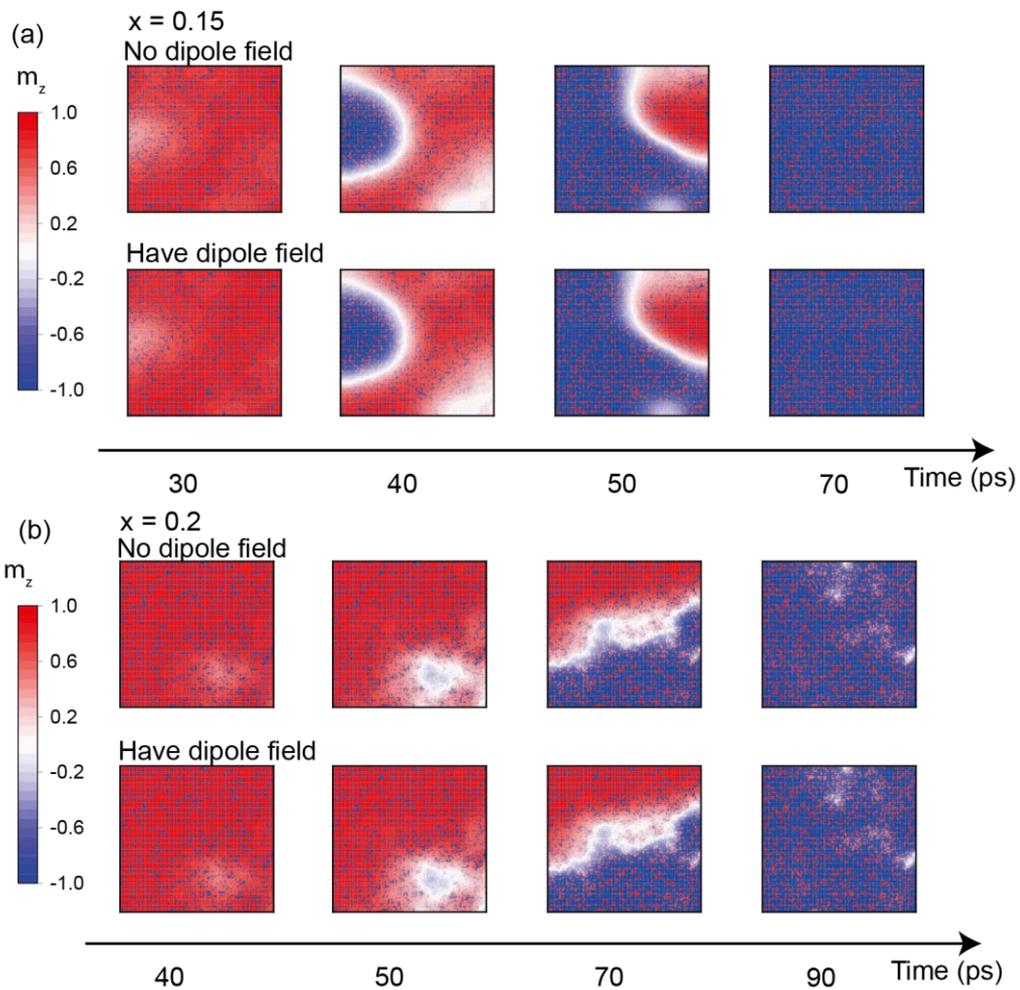

Fig. S2 The comparison of switching process with and without dipole field for the sample with (a) $x = 0.15$ and (b) $x = 0.2$.

**Note 2: Effect of FLT on the switching process**
We have simulated the STT-induced magnetization dynamics by adding field-like

torque term, which is described by the last term in the following equation:

$$\frac{\partial \mathbf{S}_i}{\partial t} = -\gamma_i \mathbf{S}_i \times \mathbf{H}_{\text{eff},i} + \alpha \mathbf{S}_i \times \frac{\partial \mathbf{S}_i}{\partial t} - \gamma_i B_{\text{D},i} \mathbf{S}_i \times (\mathbf{S}_i \times \mathbf{S}_p) + B_{\text{F},i} \mathbf{S}_i \times \mathbf{S}_p \qquad S(1),$$

where $B_{\text{D},i} = \frac{\hbar}{2} \frac{J_c \eta d^3}{\mu_i e t_{FiM}}$, $B_{\text{F},i} = \beta \cdot \frac{\hbar}{2} \frac{J_c \eta d^3}{\mu_i e t_{FiM}}$ describes the magnitudes of DLT and FLT, respectively. β represents the ratio between FLT and DLT. When β = 0.2, the result in Fig. S3 shows the same results as Fig. 2(c) (x = 0.15) in the manuscript, i.e., the first switched region tends to appear in the FeCo-rich region. Constantly increasing β, even with β = 1, the main finding in this work, i.e., the nonuniform behavior is associated with atom-rich regions, remains the same.

In addition, in previous studies, X.P. Zhao and his coworkers[s1] have verified in the experiment that the field-like torque can be neglected since it has a small value compared to the damping-like torque. Besides, in our out-of-plane system, the field-like torque would have an effective field along the z direction. It only affects the precession frequency. Therefore, we neglected this term for simplicity.

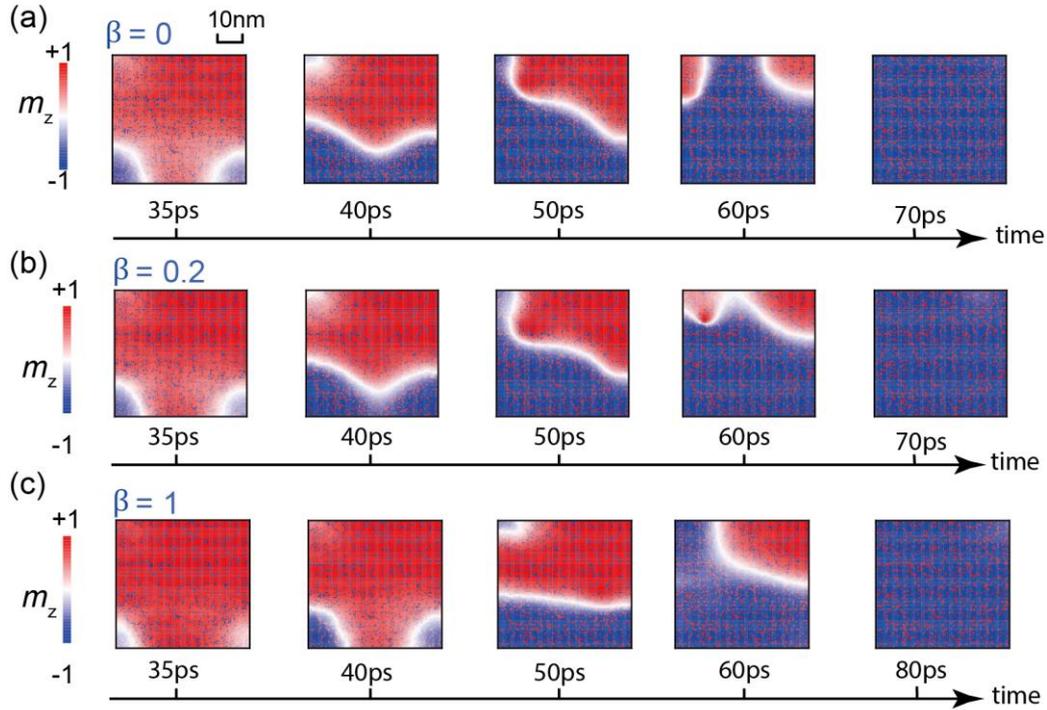

Fig. S3. The snapshot images of magnetization evolution with (a) β = 0, (b) β = 0.2, (c) β = 1.

### Note 3: Switching in the three-layer system

In the manuscript, we show the layer-resolved magnetization dynamics of a three-layer system in Fig. 1(c), and we mention that the magnetization dynamics is similar to that in layer 1. This can be better appreciated when we look at the results shown in Fig. S4, where we have prepared 3 one-layer systems that have the same atom distribution to layer 1 to 3 in the three-layer system. We then separately apply STT to the 3 one-layer system. As shown in Fig. S4, the magnetization dynamics of layer 1 is similar to that in the three-layer system, where the initial switched region appears at the right bottom corner and then propagates towards the left top corner. Therefore, we focus our discussion on the one-layer system.

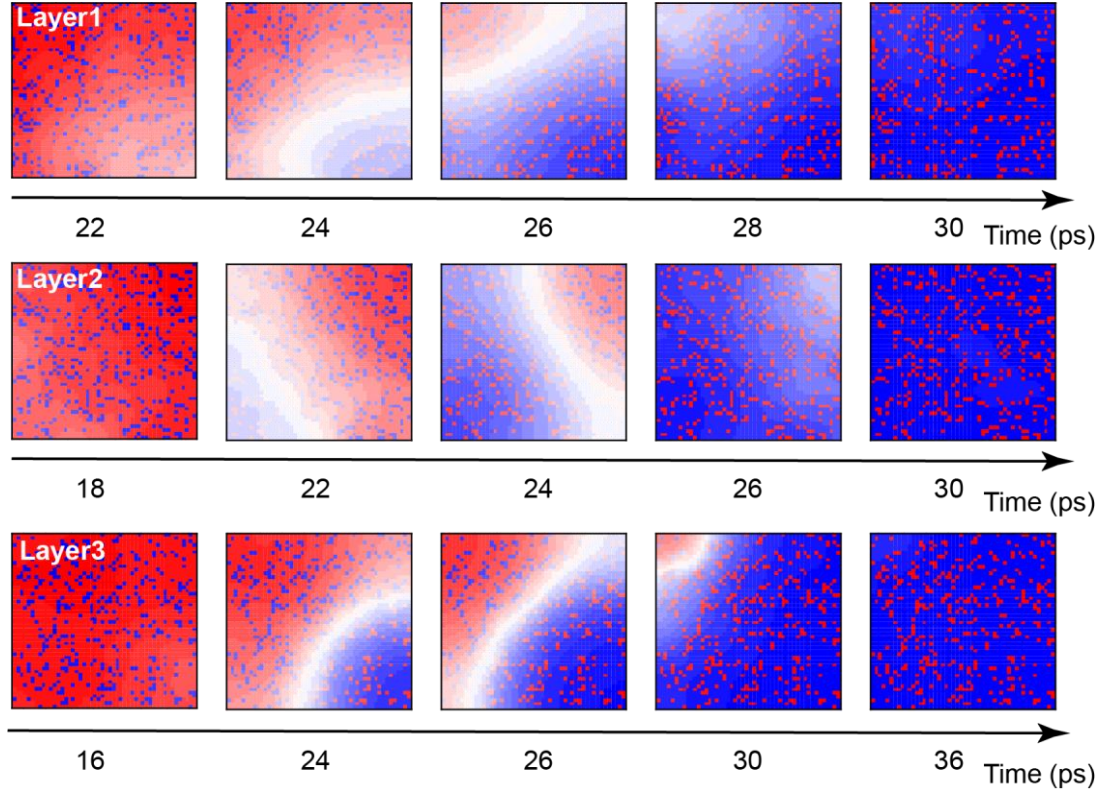

Fig. S4. The switching process of 3 one-layer system. The atom distribution of each layer is the same as the corresponding layer in the three-layer system shown in Fig. 1(c).

**Note 4: Calculation details of Fig. 3**
The atoms in our system are randomly distributed, not uniformly distributed. Taking the 2D sample with 100×100 atoms as an example, we first create a 100 ×100 grid. Then we generate a 100×100 random number matrix containing only 0 and 1, where 1 denote Gd atoms and 0 denote FeCo atoms. The numbers of 1 over 0 can be controlled in the random number matrix, which corresponds to the Gd over FeCo ratio (i.e., $x$). After that, we map the different exchange interactions ($A_{FeCo-FeCo}$ to 0-0 connection, $A_{FeCo-Gd}$ to 0-1 connection, $A_{Gd-Gd}$ to 1-1 connection) and anisotropies ($K_{FeCo}$, $K_{Gd}$).

The data in Fig. 3 is not averaged atom number over a small region. In contrast, it represents the number of neighbor Gd atoms around a center atom. Fig. S5 (a) depicts the calculation process. A 10×10 matrix is set to be the calculation matrix. This matrix works like the weight matrix in the convolutional neural network, i.e., it moves right or down one step at a time. We then calculate the number of Gd atoms in this matrix, which is then recorded in Table 2. We then use the heatmap to express the summed data. The distribution is clearly illustrated in two colors, i.e., the red region is the Gd-rich region while the blue region is the FeCo-rich region (Fig. S5(b)).

Before we decide on this representation method, we have tried other methods. For example, we first extract the atom distribution from our model as shown in Fig. S5(c) for the sample with $x = 0.5$. This distribution is nonuniform but it is difficult to visualize. Therefore, we choose to use the number of FeCo atoms in a small area to express the atom distribution of this area. We divided our sample into 100 regions. Each region contains 100 atoms. The number of FeCo atoms in each region is calculated and the results are shown in Fig. S5(d). Both Fig. S5(b) and S5(d) can express the atom-enriched regions. But we find the former method is better for visualization.

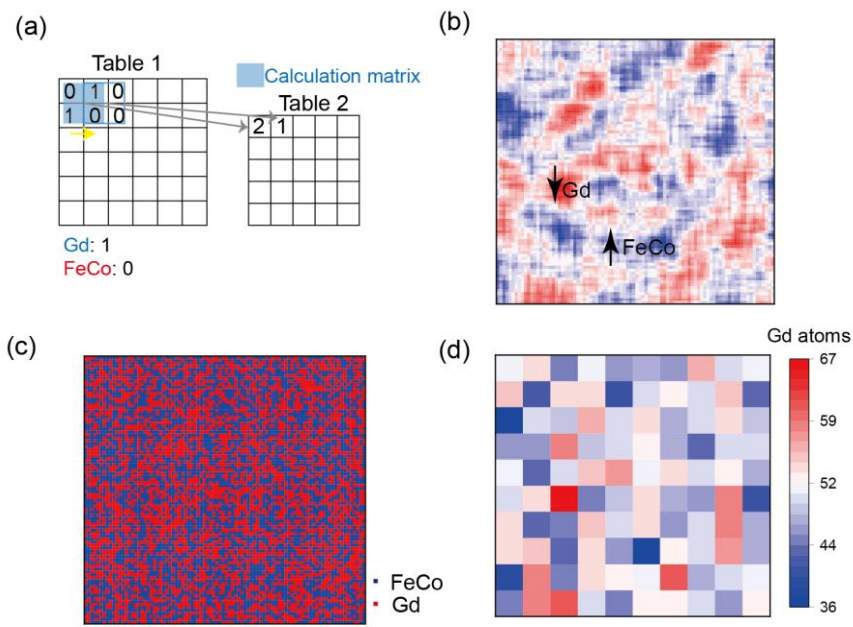

Fig. S5. (a) Illustration of the method used to describe the atom distribution. (b) (c) and (d) are the atom distribution use three different expressions, i.e., sum the around atoms, directly express, and the sum of the atoms in the small region.

**Note 5: Effect of the connected atoms on the switching dynamics**
We find that these connected atomic rich regions affect the shape of the DW. We draw the time-resolved magnetization evolution for samples with different $x$ in Fig. S6. The first figure in each row illustrates the atom distribution, where the red and blue regions represent the Gd-rich and FeCo-rich regions, respectively. The black line shows the connected Gd-rich region, i.e., the Gd cluster. When $x = 0.22$, it is clearly seen that the formation of the connected region directly affects the DW motion, e.g., the DW shape at t = 80ps is similar to the shape of the cluster. However, when the Gd-rich regions are distributed randomly without a large cluster ($x = 0.27$), the motion of the DW is flat. The Gd-rich region only affects the formation of the first switched region (t = 40ps).

We attribute these results to the difference in the exchange interaction for different atoms. For example, when $x < x_{MC}$, the FeCo-rich region is switched by STT. However, the Gd-rich and averaged region will not be switched since $\mathbf{m}_{net}$ is in the same direction as $\mathbf{S}_p$. Therefore, in these regions, the nature of the switching is the DW-induced flip through the exchange interaction between sublattices. Notably, the connected Gd clusters provide an easy path for the switching since the exchange interaction between Gd-Gd is larger than FeCo-Gd, leading to a change in the DW shape.

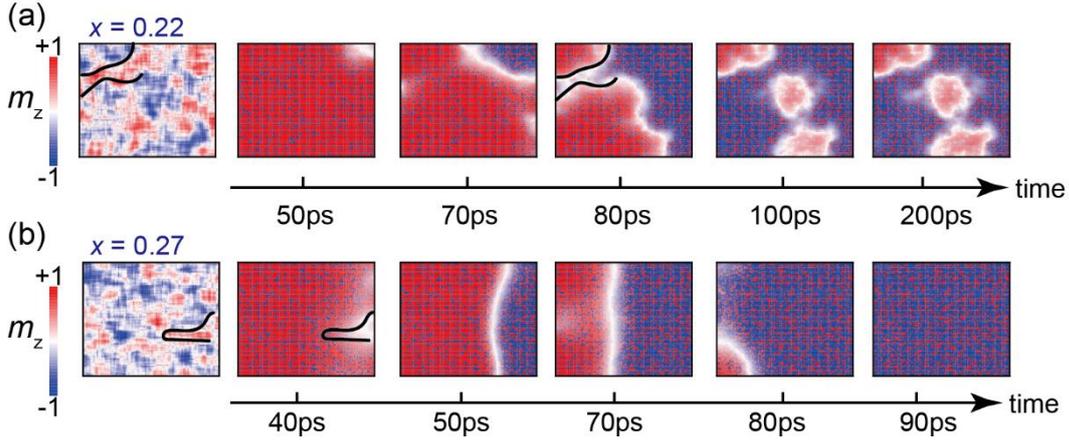

Fig. S6. The switching processes in the sample with (a) $x = 0.22$ and (b) $x = 0.27$. The first figure in each row shows the nonuniform atom distribution.

**Note 6: The Gd dominated oscillation**

In our previous work[s2], we find that in the sample with $x < x_{MC}$, the Gd atom is the oscillation-dominated sublattices and determines the oscillation direction. As shown in the figure below, independent on the current direction, the balance of damping-like STT $[-\mathbf{m} \times (\mathbf{m} \times \mathbf{S}_p)]$ and Gilbert damping torque $[-\mathbf{m} \times (\mathbf{m} \times \mathbf{H}_{eff})]$ only happens at the Gd atoms. When this torque balance is achieved, stable oscillation can be initiated. As a result, FeCo is dragged to oscillation due to the strong exchange interaction between the sublattices.

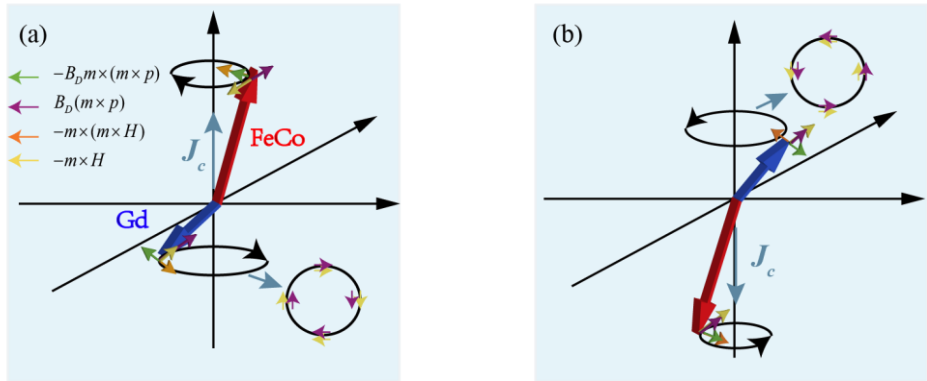

Fig. S7. Illustrations of the torques experienced by each atom when $\mathbf{m}_{FeCo}$ and $\mathbf{S}_p$ are pointing in the (a) +z and (b) −z directions.

**Note 7: Switching processes in a circular sample**
We have simulated the circle sample and we got the same result with the square sample. Fig. S8 shows the magnetization switching dynamics of the circle sample with different $x$. The results show that the shape of the sample does not affect the main results of our manuscript. Specifically, a deterministic switching occurs when $x < x_{MC}$ or $x > x_{MC}$. When $x$ approaches $x_{MC}$, a stable oscillation occurs. In addition, the magnetization evolution in the circle sample (Fig. S9) with different $x$ is also similar to the square sample, i.e., the switching process of the circle sample is nonuniform. In addition, the switching tends to occur in the FeCo-rich (Gd-rich) region when $x < x_{MC}$ ($x > x_{MC}$). In addition, the singular regions appear around the Gd-rich region.

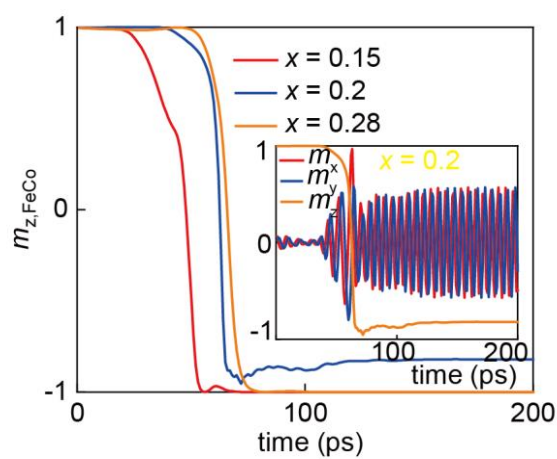

Fig. S8. Magnetization switching dynamics of the circle sample with different $x$.

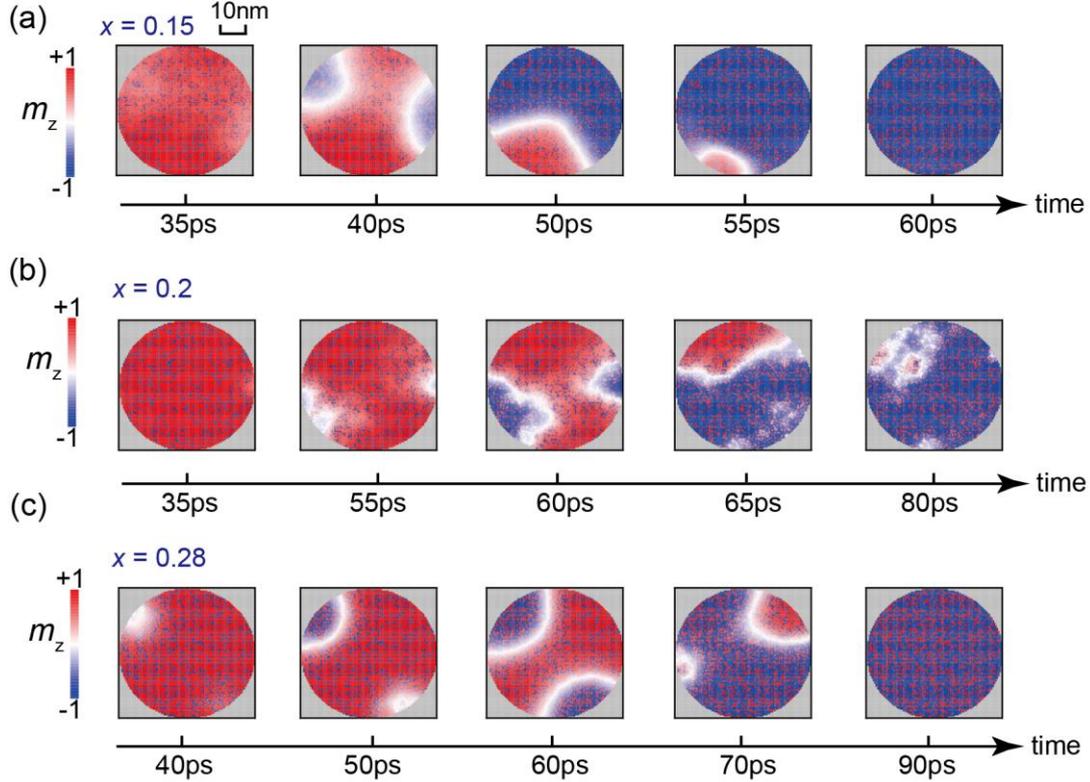

Fig. S9. Snapshot images of the switching processes in samples with different $x$.

**Note 8: P to AP and AP to P switching**

It is known that the STT efficiency η in ferromagnets (FMs) depends on the relative angle between $\mathbf{m}_{FL}$ and $\mathbf{m}_{PL}$ [s3]. However, some studies use a constant η for simplicity [s4], where they found it is sufficient to capture many key characteristics of STT switching. We find that there are not many theoretical studies of STT in FiM. The transfer of spin angular momentum might also be more complicated due to the different band diagrams of TM and RE atoms. Therefore, for the main results in our paper, we use a constant STT efficiency η for simplicity.

The angular dependent η can be expressed as $\eta(\theta) = \frac{P}{1+P^2\cos(\theta)}$, where P is the spin polarization [s5]. We use P = 0.5 [s4]. As shown in Fig. S10, in the sample with $x$ = 0.15, under the same $J_c$ = 1.1×10$^{11}$ A/m$^2$, AP to P switching is achieved, where P to AP switching does not happen, unless we increase $J_c$ to -1.2×10$^{11}$ A/m$^2$. The complete $m_z$-$J_c$ loop is shown in Fig. S10(a), where the asymmetry in $J_c$ can be observed. This can be explained by plotting η(θ) in Fig. S10(b). It shows that AP to P switching is easier due to the larger η. Based on the spatially resolved switching diagram as shown in Fig. S11(a), it is worth noting that the consideration of the angle-dependent η does not affect the switching results as we discussed in this manuscript, i.e., the initial switching region is FeCo region around the boundary when $x < x_{MC}$.

The switching results for the sample with $x$ = 0.5 are shown in Fig. S12. The first switched region is the Gd-rich region, the same as we discussed in the manuscript. However, a clear asymmetry in the $m_z$-$J_c$ loop is not observed, whereas we find that AP to P switching is faster than P to AP switching under the same $J_c$.

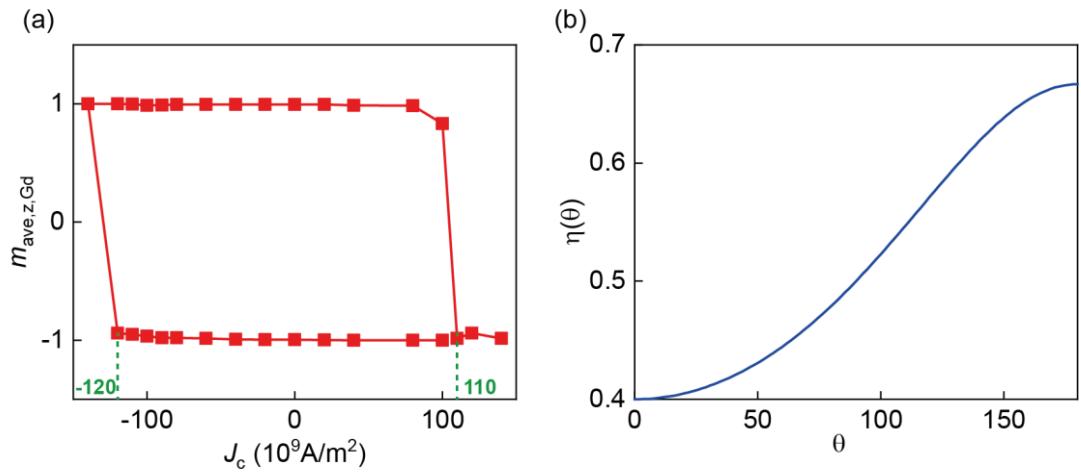

Fig. S10. (a) The $J_c$- $m_z$ loop for the sample with $x = 0.15$. Two threshold current density is $1.1\times10^{11}$ A/m$^2$ and $-1.2\times10^{11}$ A/m$^2$. (b) The relationship between the angle (between **m**$_{FL}$ and **m**$_{PL}$) and $\eta$.

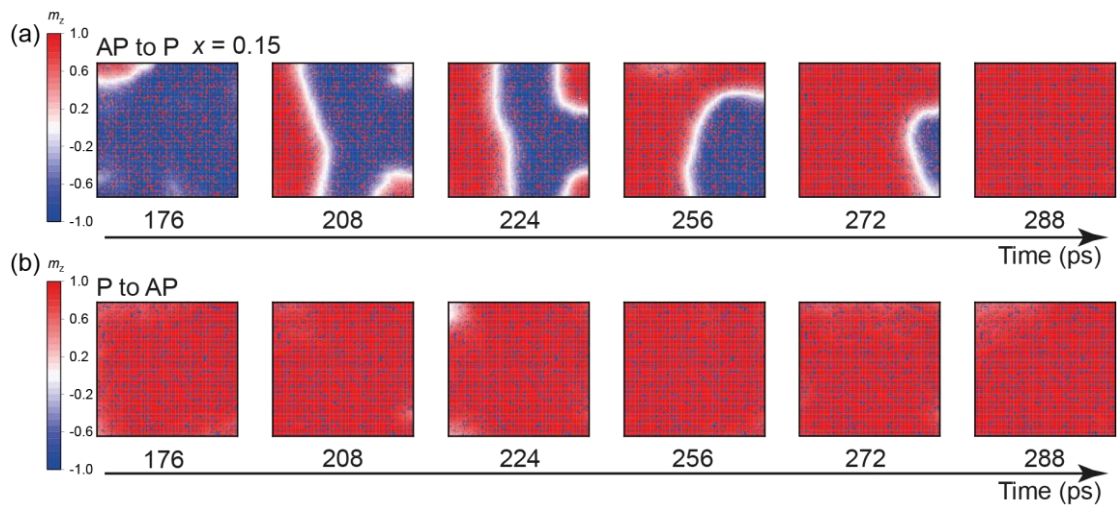

Fig. S11. The switching snapshots for the sample with $x = 0.15$. (a) AP to P switching when $J_c = 1.1\times10^{11}$ A/m$^2$. (b) P state when $J_c = -1.1\times10^{11}$ A/m$^2$.

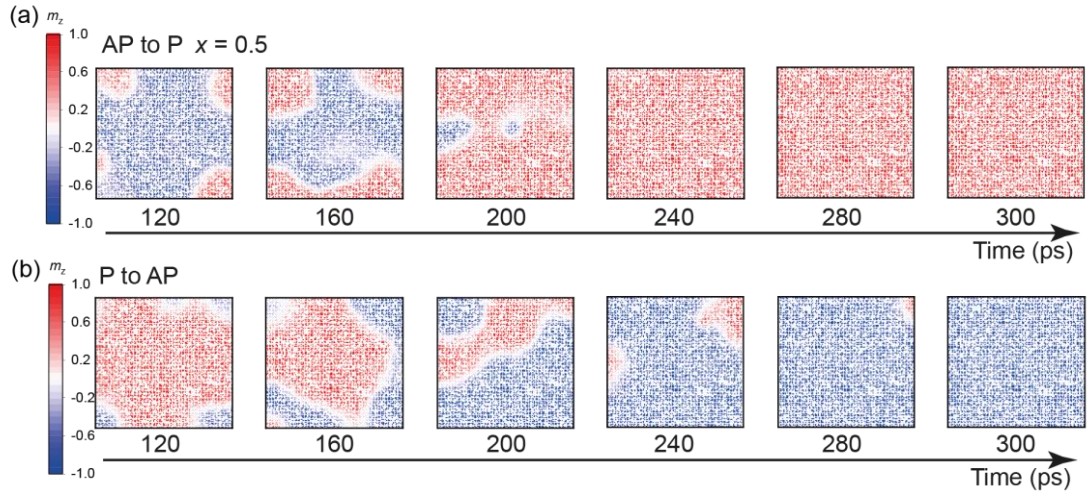

Fig. S12. The switching snapshots for the sample with $x = 0.5$. (a) AP to P switching when $J_c = 1 \times 10^{11}$ A/m$^2$. (b) P to AP switching when $J_c = -1 \times 10^{11}$ A/m$^2$.